\documentclass[sigconf,nonacm]{acmart}
\AtBeginDocument{%
  }

\setcopyright{cc}
\setcctype{by}

\usepackage{braket}
\usepackage{graphicx}
\usepackage{tikz}
\usepackage{circuitikz}

\usepackage{array}
\usepackage{arydshln} 
\usepackage{booktabs}
\usepackage{makecell}
\usepackage{amsmath}

\usepackage{enumitem}

\newcolumntype{L}[1]{>{\raggedright\arraybackslash}m{#1}}

\newcounter{individualmodel}
\begin{document}

\title{Which Superconducting Qubit Model is Good Enough?} 
\subtitle{From Effective Two-Level to Circuit-Based Hamiltonians for Pulse-Level Simulation}

\author{Frej Larssen, Ivy Peng, Stefano Markidis}
\affiliation{%
  \institution{KTH Royal Institute of Technology}
  \city{Stockholm}
  \country{Sweden}
}

\renewcommand{\shortauthors}{Larssen, Peng and Markidis}

\begin{abstract}
Pulse-level simulators are the lowest-level, most widely used abstraction layer for studying how quantum hardware responds to control signals, but they can be built on Hamiltonian models with very different fidelity and cost. This raises the question: which level of physical abstraction is sufficient for a given simulation objective? We study this question for a flux-tunable two-qubit superconducting device with a fixed bus coupler by comparing three Hamiltonian descriptions of the same hardware: an effective two-level model, a three-mode Duffing model, and a circuit-based transmon model in the charge basis. Using a realistic parameter set, we evaluate these models on a common benchmark suite spanning flux-dependent spectra, extracted two-qubit interaction terms, driven single-qubit dynamics, CZ gate dynamics, leakage outside the computational subspace, and runtime. Across the tested flux range, the Duffing model follows the circuit-based reference more closely than the effective model for static spectra and reduced two-qubit quantities, while in driven benchmarks, the multilevel models reveal effects absent in the effective description. Overall, the results support a layered use of abstraction in pulse-level simulation: effective models for reduced analyses, Duffing models as a practical multilevel default, and circuit-based models for high-fidelity reference simulation or detailed leakage analysis.
\end{abstract}




\maketitle
\section{Introduction}
Pulse-level simulation is the lowest abstraction level commonly used in quantum hardware simulation before one reaches full device-specific electromagnetic or circuit-design workflows. Instead of representing a quantum program as a sequence of ideal gates, it represents the hardware as a controlled dynamical system and simulates how externally applied signals drive its quantum state over time~\cite{majidy2024building,krantz2019quantum,qiskit_dynamics_docs,qiskit_dynamics_backend_tutorial}. For this reason, pulse-level simulation is the natural setting for high-fidelity studies of quantum devices, as it retains the hardware-level dynamics from which gate operations arise.

In a pulse-level simulation, the dynamics are governed by a time-dependent Hamiltonian, determined by both the internal structure of the device and the applied controls. A pulse-level simulator does two main kinds of computations. For fixed control settings, it computes eigenvalues and eigenstates of the Hamiltonian in order to obtain energy levels, transition frequencies (the frequencies associated with transitions between energy levels), and dressed states of the interacting system (eigenstates of the coupled qubit-coupler system rather than of the uncoupled subsystems). For time-dependent controls, it propagates the quantum state under the Hamiltonian to predict the dynamical response of the device during a pulse.

In this work, we focus on superconducting qubits, where pulse-level simulation is especially important because control is implemented directly through analog signals. In superconducting circuits, the Hamiltonian is modified through two main control channels. The first is a \emph{drive pulse}, typically a microwave control signal, which acts directly on a qubit and is used mainly for single-qubit gates. The second is a \emph{flux pulse}, or time-dependent flux-bias control, which changes qubit frequencies and therefore changes the interaction pattern of the device. In this work, the single-qubit benchmark uses a drive pulse and the two-qubit benchmark uses a flux pulse. 

The basic components of the device are qubits, couplers, and control channels. The qubits store the computational states, the coupler mediates interaction between them, and the control signals modify the Hamiltonian over time. Superconducting circuits are therefore a natural setting for pulse-level simulation. The same logical operation may arise from underlying dynamics that can be modeled at several levels of physical detail. The central question is then one of abstraction: \emph{which Hamiltonian description is sufficient for the simulation objective?} A reduced model may be fast enough for repeated sweeps, fitting, or control optimization, but it may omit dynamical effects that become important under realistic control pulses~\cite{chen2016leakage,negirneac2021cz}. A more detailed model may capture these effects, but at a substantially higher numerical cost.

We evaluate the models using a benchmark suite that includes both static and time-domain tasks. The static benchmarks test how well each model reproduces dressed spectra and reduced two-qubit quantities as the controls vary across operating points. The time-domain benchmarks solve the driven dynamics under explicit pulses and test single-qubit behavior, CZ gate dynamics, and leakage outside the computational subspace. We also measure runtime. 

The main contributions of this paper are:
\begin{itemize}
    \item We compare three Hamiltonian descriptions for pulse-level simulation of a superconducting two-qubit device with a fixed harmonic bus: an effective two-level model, a three-mode Duffing model, and a circuit-based transmon model in the charge basis.
    \item We evaluate these models on a common set of benchmarks covering both static and dynamical behavior, including dressed spectra, extracted effective quantities, driven single-qubit dynamics, CZ gate dynamics, leakage outside the computational subspace, and computational cost.
    \item We show that the ranking of the models depends on the observable: the Duffing model is more faithful for flux-dependent static structure, the effective model can still be competitive for reduced observables such as the CZ conditional phase, and the circuit-based model is needed when detailed leakage pathways must be resolved.
    \item We provide practical guidance on when each model is sufficient, ranging from rapid reduced analysis to leakage-sensitive pulse-level simulation.
\end{itemize}

\section{Preliminaries}
In gate-level simulation, a quantum program is represented as a sequence of ideal unitary gates acting on a finite-dimensional computational state space. Pulse-level simulation uses a lower-level description. Instead of assuming ideal gates, it models the physical device directly through a Hamiltonian that depends on externally applied controls such as microwave drives and flux biases~\cite{krantz2019quantum,qiskit_dynamics_docs,puzzuoli2023qiskitdynamics}. 

In pulse-level simulations, the Hamiltonian is the central quantity. It specifies both the internal structure of the device and how control signals act on it. The quantum state $\ket{\psi(t)}$ represents the instantaneous state of the full physical system. Depending on the model, that state may include only the computational states of the qubits, or it may also include higher excited qubit levels and excitations of the coupler. The time evolution is governed by the time-dependent Schr\"odinger equation
\begin{equation}
i\frac{d}{dt}\ket{\psi(t)} = H(t)\ket{\psi(t)}.
\label{eq:tdse_background}
\end{equation}
Solving this equation leads to the state trajectory induced by the controls. In this sense, pulse-level simulation treats the device as a controlled continuous-time dynamical system. A gate can be viewed as a compressed description of what the system does over a finite time window. If the control pulse from time $t_0$ to $t_1$ induces the unitary
\begin{equation}
U(t_1,t_0)=\mathcal{T}\exp\!\left(-i\int_{t_0}^{t_1} H(t)\,dt\right),
\label{eq:time_ordered_unitary}
\end{equation}
where $\mathcal{T}$ denotes time ordering, then the gate-level description keeps only $U(t_1,t_0)$, not the intermediate state trajectory during that interval. In this sense, the gate-level view is a stroboscopic view of the underlying pulse-level dynamics~\cite{majidy2024building}. In this work, we focus on transmon qubits. For a thorough review on superconducting qubits such as transmons, see Refs.~\cite{koch2007charge, krantz2019quantum}.

An important issue in pulse-level simulations is the representation of the state space (the set of states that the simulator allows the system to occupy). The physical models considered in this work are not naturally finite-dimensional. Oscillator-based models are infinite-dimensional because an oscillator can, in principle, occupy arbitrarily many excitation levels. Likewise, circuit-based transmon models written in the \emph{charge basis} represent the qubit state as a superposition of charge-number states, where each basis state corresponds to a different integer number of excess Cooper pairs on the superconducting island. Since that charge number is not bounded a priori, the corresponding basis is also infinite in principle~\cite{koch2007charge,krantz2019quantum}. From a computational point of view, truncation is part of the model definition itself. It determines the dimension of the matrices used in eigendecomposition and time propagation. Therefore, it directly affects both runtime and memory cost. It also affects physical fidelity, since omitted states can no longer participate in the simulated dynamics.

\section{System, models, and comparison protocol}

\noindent \textbf{Device architecture.} We study an interacting quantum system consisting of three subsystems: two flux-tunable transmon qubits and one shared harmonic bus. From a simulation point of view, this architecture is useful because it is simple enough to analyze across multiple Hamiltonian descriptions, while still exhibiting the main effects that matter in pulse-level simulation. The qubits are the computational subsystems, while the bus acts as an intermediate mode that couples them and allows energy to be exchanged between them.

One of the control inputs is the applied flux. Here, \emph{flux} means the externally applied magnetic flux threading the SQUID loop of a split-junction transmon. Because the effective Josephson energy depends on that flux, the qubit frequency also becomes flux dependent. This is the sense in which the qubits are \emph{flux tunable}. Changing the external flux changes the qubit frequency, and therefore alters how strongly the qubit interacts with the bus and with the other qubit, for example through resonance.

The other control input is the microwave drive, realized as an alternating voltage or current applied through a capacitively coupled line to each transmon~\cite{krantz2019quantum}. When the drive frequency is close to the qubit transition frequency, energy can be transferred to or from the transmon.

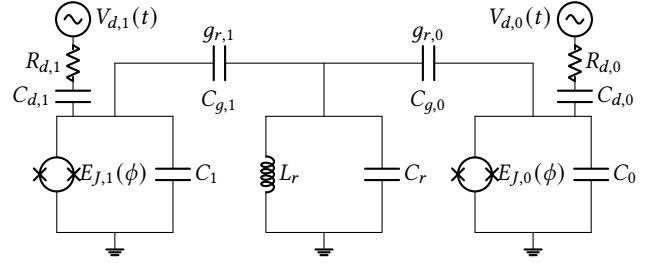
\begin{figure}[t]
\pgfmathsetmacro{\BusMarginY}{20.0}
\pgfmathsetmacro{\DriveCapMarginY}{15.0}
\pgfmathsetmacro{\DriveResMarginY}{25.0}
\pgfmathsetmacro{\DriveVolMarginY}{5.0}
\pgfmathsetmacro{\DriveStubMarginY}{13.0}
\pgfmathsetmacro{\BottomY}{0.00}

\pgfmathsetmacro{\DistToWidthRatio}{0.8}
\pgfmathsetlengthmacro{\LeftX}{0pt}
\pgfmathsetlengthmacro{\RightX}{\linewidth-40} 
\pgfmathsetlengthmacro{\TransmonWidth}{\RightX / (3 + 2*\DistToWidthRatio)}
\pgfmathsetlengthmacro{\TopY}{\TransmonWidth}
\pgfmathsetlengthmacro{\BusY}{\TopY+\BusMarginY}
\pgfmathsetlengthmacro{\DriveCapY}{\TopY+\DriveCapMarginY}
\pgfmathsetlengthmacro{\DriveResY}{\TopY+\DriveResMarginY}
\pgfmathsetlengthmacro{\DriveY}{\DriveResY+\DriveVolMarginY}
\pgfmathsetlengthmacro{\DriveStubY}{\DriveY+\DriveStubMarginY}
\pgfmathsetlengthmacro{\TransmonDist}{\DistToWidthRatio * \TransmonWidth}
\pgfmathsetlengthmacro{\LeftCapX}{\LeftX + \TransmonWidth}
\pgfmathsetlengthmacro{\ResLeftX}{\LeftCapX + \TransmonDist}
\pgfmathsetlengthmacro{\ResRightX}{\ResLeftX + \TransmonWidth}
\pgfmathsetlengthmacro{\RightCapX}{\RightX - \TransmonWidth}
\pgfmathsetlengthmacro{\LeftTapX}{0.5*(\LeftX + \LeftCapX)}
\pgfmathsetlengthmacro{\ResCenterX}{0.5*(\ResLeftX + \ResRightX)}
\pgfmathsetlengthmacro{\RightTapX}{0.5*(\RightCapX + \RightX)}
\pgfmathsetlengthmacro{\DriveOffset}{0.45*\TransmonDist}
\pgfmathsetlengthmacro{\LeftDriveX}{\LeftTapX - \DriveOffset}
\pgfmathsetlengthmacro{\RightDriveX}{\RightTapX + \DriveOffset}

\begin{circuitikz}[line cap=round, line join=round]
  \ctikzset{
    bipoles/length=0.75cm,
    capacitors/scale=1.0,
    resistors/scale=0.75,
    inductors/scale=1.0
  }

  \draw
    (\LeftTapX,\BusY)
    to[C, l_=$C_{g,1}$, a^=$g_{r,1}$] (\ResCenterX,\BusY)
    to[C, l_=$C_{g,0}$, a^=$g_{r,0}$] (\RightTapX,\BusY);

  \draw (\LeftX,\TopY) -- (\LeftCapX,\TopY);
  \draw (\LeftTapX,\TopY) -- (\LeftTapX,\BusY);
  \draw (\LeftDriveX,\TopY)
    to[C, l=$C_{d,1}$] (\LeftDriveX,\DriveCapY)
    to[R, l=$R_{d,1}$] (\LeftDriveX,\DriveResY)
    -- (\LeftDriveX,\DriveY)
    to[sV, l_=$V_{d,1}(t)$] (\LeftDriveX,\DriveStubY);
  \draw (\LeftX,\TopY) to[squid,l=$E_{J,1}(\phi)$] (\LeftX,\BottomY);
  \draw (\LeftCapX,\TopY) to[C,l=$C_1$] (\LeftCapX,\BottomY);
  \draw (\LeftX,\BottomY) -- (\LeftCapX,\BottomY);
  \draw (\LeftTapX,\BottomY) node[ground] {};

  \draw
    (\ResLeftX,\TopY)
    to[cute inductor,l=$L_r$] (\ResLeftX,\BottomY)
    -- (\ResRightX,\BottomY)
    to[C,l_=$C_r$] (\ResRightX,\TopY)
    -- (\ResLeftX,\TopY);
  \draw (\ResCenterX,\TopY) -- (\ResCenterX,\BusY);
  \draw (\ResCenterX,\BottomY) node[ground] {};

  \draw (\RightCapX,\TopY) -- (\RightX,\TopY);
  \draw (\RightTapX,\TopY) -- (\RightTapX,\BusY);
  \draw (\RightDriveX,\TopY)
    to[C, l_=$C_{d,0}$] (\RightDriveX,\DriveCapY)
    to[R, l_=$R_{d,0}$] (\RightDriveX,\DriveResY)
    -- (\RightDriveX,\DriveY)
    to[sV, l=$V_{d,0}(t)$] (\RightDriveX,\DriveStubY);
  \draw (\RightCapX,\TopY) to[squid,l=$E_{J,0}(\phi)$] (\RightCapX,\BottomY);
  \draw (\RightX,\TopY) to[C,l=$C_0$] (\RightX,\BottomY);
  \draw (\RightCapX,\BottomY) -- (\RightX,\BottomY);
  \draw (\RightTapX,\BottomY) node[ground] {};
\end{circuitikz}
  \Description{Superconducting circuit drawing.}
 \caption{Two flux-tunable transmon qubits ($q_0$ on the right and $q_1$ on the left) are coupled through a fixed harmonic bus, represented as an intermediate LC resonator. The subsystem ordering is $\ket{q_1,c,q_0}$, with $q_0$ the least significant qubit. Each transmon is coupled to a microwave drive line, modeled here by the time-dependent voltage $V_{d,j}(t)$.}
  \label{fig:superconducting-circuit-drawing}
\end{figure}

Figure~\ref{fig:superconducting-circuit-drawing} shows the circuit studied in this work. Each qubit is implemented as a flux-tunable transmon, and the two qubits are coupled through a fixed harmonic bus represented as an LC mode. This gives a three-mode device with one shared coupler and two controlled qubit modes. Each transmon is coupled to a microwave drive line. Related bus-mediated couplings have been used since the early generations of superconducting processors and remain relevant in current two-qubit gate designs~\cite{majer_coupling_2007,dicarlo_demonstration_2009,dicarlo_preparation_2010,jiang2025microwave}. \\

\noindent \textbf{Hamiltonian decomposition and studied models.} At pulse level, the simulated object is a time-dependent Hamiltonian acting on the state space of the device. We decompose it as
\begin{equation}
H(t)=\sum_{j\in\{0,1\}} H_{q,j}(\phi_j)
      + H_c
      + H_i
      + \sum_{j\in\{0,1\}} H_{d,j}(t),
\label{eq:full_decomposition}
\end{equation}
where $H_{q,j}$ describes qubit $j$, $H_c$ describes the shared bus mode, $H_i$ contains the couplings between subsystems, and $H_{d,j}(t)$ is the time-dependent drive applied to qubit $j$.

The three models studied in this paper differ primarily in the state space on which Eq.~\eqref{eq:full_decomposition} acts. The effective two-level model keeps only the four computational basis states of the two qubits. The three-mode Duffing model retains explicit multilevel state spaces for the two qubits and the bus, but approximates the qubits as weakly anharmonic oscillators. The circuit-based transmon model starts from a larger charge-basis description of each qubit and therefore stays closest to the underlying circuit parameters. In this work we only consider the flux dependence on one of the qubits. For the rest of this paper, $\phi$ is shorthand for $\phi_1$, and we fix $\phi_0=0$. \\[-3pt]

\noindent \textit{Effective two-level model.}
We call this model \emph{effective} because it does not represent all physical degrees of freedom explicitly. Instead, it keeps only the four computational basis states of the two qubits and replaces the omitted multilevel and bus dynamics by a small set of fitted parameters. In other words, the coupler and higher excited states are not simulated directly. Their influence is compressed into renormalized qubit frequencies and effective two-qubit couplings. This produces a reduced model that is intended to reproduce the behavior seen inside the computational subspace without resolving the full underlying dynamics.

Therefore, the model acts only on the computational subspace of the two qubits, so its state space has dimension four. Its drift Hamiltonian is
\begin{equation}
H_{\mathrm{eff}}(\phi)
=
\sum_{j\in\{0,1\}} \frac{\tilde{\omega}_j(\phi)}{2} Z_j
+J(\phi)\bigl(X_1X_0+Y_1Y_0\bigr)
+\frac{\zeta(\phi)}{4} Z_1 Z_0,
\label{eq:effective_model}
\end{equation}
where $X_j$, $Y_j$, and $Z_j$ are Pauli operators on qubit $j$, i.e., the standard $2\times 2$ matrices used to describe single-qubit transformations in the computational basis. Here, $\tilde{\omega}_j(\phi)$ are dressed qubit frequencies, $J(\phi)$ is an exchange-like coupling between the two qubits, and $\zeta(\phi)$ is the residual $ZZ$ interaction. Intuitively, $J(\phi)$ quantifies how strongly the qubits can exchange excitation, while $\zeta(\phi)$ measures how much the energy of one qubit depends on the state of the other, which underlies conditional phase accumulation. In this work, these quantities are extracted from the circuit-based reference model projected onto the computational subspace, and then fitted as functions of flux. \\[-3pt]

\noindent \textit{Three-mode Duffing model.}
The Duffing model is an intermediate description between the four-state effective model and the more detailed circuit-based model. It keeps the two qubits and the bus as explicit dynamical modes, so population can move not only within the computational basis but also into higher excited levels of the qubits and the bus. At the same time, it does not retain the full circuit variables of the transmons. Instead, each qubit is approximated as a weakly anharmonic oscillator.

In this model, the qubit is modeled as an oscillator with a small nonlinear correction to the harmonic spectrum. A purely harmonic oscillator has evenly spaced energy levels, which is not suitable for modeling a qubit since driving one transition would also excite others at the same frequency. The Duffing correction introduces a weak anharmonicity, so the level spacing becomes slightly uneven. This makes the lowest transition distinguishable from higher ones and provides a compact approximation to transmon dynamics while keeping the model numerically manageable.

Its drift Hamiltonian is
\begin{equation}
\begin{aligned}
H_{\mathrm{Duff}}(\phi)
={}&
\sum_{j\in\{0,1\}}
\left(
\omega_j(\phi)\, a_j^\dagger a_j
+\frac{\alpha_j}{2} a_j^\dagger a_j^\dagger a_j a_j
\right)
+\omega_c\, a_c^\dagger a_c \\
&\quad
+\sum_{j\in\{0,1\}} g_{j,c}\left(a_j^\dagger a_c+a_c^\dagger a_j\right),
\end{aligned}
\label{eq:duffing_model}
\end{equation}
where $a_j$ and $a_c$ are annihilation operators for qubit $j$ and the bus, $\omega_j(\phi)$ and $\omega_c$ are their mode frequencies, $\alpha_j$ is the anharmonicity of qubit $j$, and $g_{j,c}$ is the qubit-bus coupling strength. The annihilation operators $a_j$ and $a_c$ lower the excitation number of qubit mode $j$ and of the bus mode, respectively. They are the standard operators used in oscillator models to represent how energy is stored, exchanged, and driven in each mode. Intuitively, this model treats the device as three coupled oscillatory modes, but with the qubit modes made slightly nonuniform so that they behave like qubits rather than ideal resonators. \\[-3pt]

\noindent \textit{Circuit-based transmon model.} The circuit-based model is the most detailed representation considered in this paper. It starts from a larger \emph{charge basis}, in which the qubit state is expanded over charge-number states. This keeps a closer connection to the underlying circuit variables before truncation and therefore preserves more of the device structure than the effective or Duffing descriptions.

Its qubit Hamiltonian is
\begin{equation}
H_{q,\mathrm{circ}}=
\sum_{j\in\{0,1\}}
\left[
4E_{C,j}(\hat n_j-n_{g,j})^2
-\frac{E_{J,j}(\phi_j)}{2}
\sum_n \left(\ket{n}\bra{n+1}+\mathrm{h.c.}\right)
\right],
\label{eq:circuit_qubit_model}
\end{equation}
where $E_{C,j}$ and $E_{J,j}(\phi_j)$ are the charging and Josephson energies of qubit $j$, $\hat n_j$ is the charge-number operator,  $n_{g,j}$ is the offset charge, and  $\mathrm{h.c.}$ is the Hermitian conjugate. The first term represents charging energy in the charge basis, while the second term couples neighboring charge states through Josephson tunneling. Together, these terms define the transmon as a circuit-level quantum system rather than as a reduced oscillator approximation.

The coupler is represented by the same harmonic term $\omega_c a_c^\dagger a_c$ as in the Duffing model, while the interaction is
\begin{equation}
H_{i,\mathrm{circ}}=
\sum_{j\in\{0,1\}} g_j \hat n_j \left(a_c+a_c^\dagger\right).
\label{eq:circuit_interaction_model}
\end{equation}
This means that the qubit-bus interaction is expressed directly in terms of the qubit charge operator and the bus quadrature, instead of through an already reduced effective coupling.

The Josephson energy of qubit $j$ is tuned by flux according to
\begin{equation}
E_{J,j}(\phi_j)=E_{J,j}^{\max}
\sqrt{\cos^2(\pi\phi_j)+d_j^2\sin^2(\pi\phi_j)},
\label{eq:ej_flux_dependence}
\end{equation}
where $\phi_j=\Phi_j/\Phi_0$ is the reduced applied flux and $\Phi_0$ is the flux quantum. Through this dependence, the external flux changes the qubit frequency and therefore changes the interaction pattern of the full device. \\

\noindent \textit{Modeling of the drive.} We model single-qubit control by the drive term
\begin{equation}
  H_{d,q} = -\frac{1}{2}A \left(e^{-i\theta} a^\dagger + e^{i\theta} a\right),
\end{equation}
where $A$ is the microwave amplitude, $\theta$ is the drive phase, and $a$ is the lowering operator. The difference between the effective, Duffing, and circuit models is how $a$ is realized. In the effective model, $a=\lvert 0\rangle\langle 1\rvert$. In the Duffing and circuit-based models it is represented by the corresponding lowering operator on the full multilevel state space.\\

\noindent\textbf{Comparison protocol, benchmarks, and parameter setting.}
Table~\ref{tab:system_parameters} lists the device parameters used in the benchmarks. The physical parameters are chosen to mimic a fixed-bus two-qubit superconducting device similar to that of Ref.~\cite{dicarlo_demonstration_2009}. The regime $E_{J,j}^{\max}\gg E_{C,j}$ is consistent with standard transmon operation~\cite{koch2007charge,krantz2019quantum}. In this regime, the qubit becomes much less sensitive to charge noise while still retaining enough anharmonicity to function as a controllable qubit rather than as a purely harmonic oscillator. All energies are reported in GHz.

\begin{table}[ht!]
\centering
\small
\setlength{\tabcolsep}{4pt}
\renewcommand{\arraystretch}{1.05}
\begin{tabular}{@{}llcc@{}}
\toprule
\textbf{Category} & \textbf{Parameter} & \textbf{$q_1$} & \textbf{$q_0$ / shared} \\
\midrule
Qubit       & $E_{J,\max}$       & $28.48~\mathrm{GHz}$ & $42.34~\mathrm{GHz}$ \\
Qubit       & $E_C$              & $0.317~\mathrm{GHz}$ & $0.297~\mathrm{GHz}$ \\
Coupler     & $E_{\mathrm{osc}}$ & \multicolumn{2}{c}{$6.902~\mathrm{GHz}$} \\
Coupler     & $\kappa/2\pi$      & \multicolumn{2}{c}{$0.001~\mathrm{GHz}$} \\
Interaction & $g_{j,c}$           & $0.183~\mathrm{GHz}$ & $0.199~\mathrm{GHz}$ \\
\bottomrule
\end{tabular}
\caption{System parameters used in the benchmarks.}
\label{tab:system_parameters}
\end{table}
In all benchmarks, the circuit-based reference model is implemented in \texttt{scqubits} using two tunable transmons and one harmonic oscillator~\cite{groszkowski2021scqubits}. Each qubit is first represented in a charge basis of dimension $N_Q$, then projected to its $N_{E,q}$ lowest-energy eigenstates. The coupler is truncated to $N_{E,c}$ harmonic levels. These truncations define the finite-dimensional state space on which the simulations are carried out.

The reduced models are calibrated using only static observables extracted from the circuit-based reference model. All time-domain results are out-of-sample tests of model transfer. In other words, the reduced Hamiltonians are fitted to match fixed-point spectral structure, and are then assessed on their ability to generalize to driven dynamics that were not used during calibration.

For the effective model, the fitted quantities are derived from the dressed spectrum of the circuit model. In particular, the dressed qubit frequencies are obtained from
\begin{equation}
\tilde{\omega}_j(\phi)
=
E_{\ket{\mathsf{1}_j}}(\phi)-E_{\ket{00}}(\phi)+\frac{1}{2}\zeta(\phi),\quad\ket{\mathsf{1}_0} = \ket{01}, \ket{\mathsf{1}_1} = \ket{10}
\label{eq:wtilde_target}
\end{equation}
and are represented by a low-order trigonometric fit. Here, $E_{\lvert \mathsf{1}_j\rangle}$ denotes the dressed energy of the computational state in which qubit $j$ is excited and the other qubit is in $\lvert 0\rangle$. This fit is natural here because the flux dependence is smooth and approximately periodic, so a small number of harmonics captures the dominant variation while keeping the reduced model compact and easy to evaluate. The flux dependence of $J(\phi)$ and $\zeta(\phi)$ is then fitted separately using regularized inverse-detuning surrogate functions, capturing the peaks of exchange near resonance with the coupler.

For the Duffing model, the qubit frequency and anharmonicity are initialized by a least-squares fit to the circuit parameters using a trigonometric fit. The bus frequency $\omega_c$ and couplings $g_{j,c}$ are taken directly from the device parameters.

We compare the three models with the following benchmarks:
\begin{enumerate}[label=\emph{\roman*)}, leftmargin=0pt, itemindent=18pt, labelsep=6pt, itemsep=2pt, topsep=2pt, parsep=0pt]
\item \emph{Static spectra across flux.}
For each fixed flux value, we construct the corresponding time-independent Hamiltonian and compute its eigendecomposition. The resulting eigenvalues define the dressed energy levels of the interacting system at that operating point. Comparing these spectra across models tests whether a given Hamiltonian description preserves the low-energy structure of the device and how that structure changes as the control parameter is varied.

\item \emph{Extracted effective two-qubit quantities.}
From the dressed computational manifold, we extract the residual interaction
\begin{equation}
\zeta(\phi)=E_{00}(\phi)-E_{01}(\phi)-E_{10}(\phi)+E_{11}(\phi),
\label{eq:zeta_def}
\end{equation}
and the exchange-like coupling
\begin{equation}
J(\phi)=\frac{1}{2}\mathrm{Re}\!\left[\bra{01}H_{\mathrm{eff}}(\phi)\ket{10}\right],
\label{eq:J_def}
\end{equation}
where $H_{\mathrm{eff}}(\phi)$ is the effective $4\times 4$ Hamiltonian on the computational subspace. Here, $E_{00}$, $E_{01}$, $E_{10}$, and $E_{11}$ are the dressed energies associated with computational basis states $\ket{q_1 q_0}=\ket{00}, \ket{01}, \ket{10},$ $ \ket{11}$.

\item \emph{Driven single-qubit dynamics.}
We next study an $R_X$-type benchmark in which one qubit is driven while the other remains undriven. The undriven qubit is the \emph{spectator qubit}. It is not directly controlled, but it remains coupled to the system and may influence the dynamics. We repeat the same pulse with the spectator initialized in both $\ket{0}$ and $\ket{1}$. This benchmark probes three aspects of the response. \emph{Population transfer} measures how much probability moves between the intended computational basis states under the pulse. \emph{Spectator dependence} measures how much that response changes with the initial state of the undriven qubit. \emph{Leakage} measures probability transfer from the computational subspace into higher excited states outside the ideal two-level qubit model.

\item \emph{CZ gate dynamics.}
We then study conditional-phase accumulation generated by a flux pulse on one qubit. Conditional-phase accumulation means that the joint two-qubit state acquires a phase that depends on the computational state of both qubits, rather than only on each qubit individually. This is the mechanism underlying CZ-type gates. One basis state, typically $\ket{11}$, accumulates an extra relative phase with respect to the others. 
The pulse is chosen using the static benchmark, namely by moving to a flux region where the extracted $ZZ$ interaction is enhanced. The resulting conditional phase is measured as
\begin{equation}
\phi_{\mathrm{CZ}}(t)=\phi_{11}(t)+\phi_{00}(t)-\phi_{01}(t)-\phi_{10}(t),
\label{eq:conditional_phase}
\end{equation}
where $\phi_{ab}(t)$ denotes the accumulated phase of the computational basis state $\ket{ab}$ under the driven evolution.

\item \emph{Leakage.}
Finally, we quantify population transfer outside the computational subspace
\begin{equation}
\mathcal{H}_{\mathrm{comp}}=\mathrm{span}\{\ket{00},\ket{01},\ket{10},\ket{11}\},
\label{eq:background_hcomp}
\end{equation}
during the CZ-evolution.
If $\Pi_{\mathrm{comp}}$ is the projector onto $\mathcal{H}_{\mathrm{comp}}$, then the leakage predicted by model $m$ at time $t$ is
\begin{equation}
L^{(m)}(t)=
1-\bra{\psi^{(m)}(t)}\Pi_{\mathrm{comp}}\ket{\psi^{(m)}(t)}.
\label{eq:leakage_def}
\end{equation}
Leakage is particularly informative in this comparison because the effective two-level model cannot represent it explicitly, whereas the Duffing and circuit-based models contain non-computational levels by construction.
\end{enumerate}

Whenever dressed energies are mapped back to computational states, the assignment is based on overlap with the corresponding bare basis states. We use local overlap-based matching at each flux point.

In addition to the physical observables above, we also measure numerical cost. Specifically, we record the time required to construct the Hamiltonian objects and the time required to propagate the state under a prescribed flux pulse. This allows us to compare the models not only in terms of which observables they reproduce, but also in terms of the computational effort required to use them in practice. \\

\noindent\textbf{Numerical solution procedure.}
All reported quantities are obtained from finite-dimensional approximations of the model Hamiltonians. For static benchmarks, we fix the control parameter $\phi$ and construct the corresponding time-independent circuit Hamiltonian in \texttt{scqubits}. Each transmon is first represented in charge basis of dimension $N_Q$ and then diagonalized with \texttt{SciPy}, retaining the lowest $N_{E,q}$ qubit eigenstates. The coupler is represented in a truncated oscillator basis, and the full Hamiltonian is assembled by extending subsystem operators with identities and adding the interaction terms.\footnote{Map between \texttt{scqubits} notation and this paper: $N_Q = 2\,\mathtt{ncut}+1$ and $N_E=\mathtt{truncated\_dim}$.}

Some of the quantities used in the comparison are defined on a two-qubit computational subspace. For this reason, we construct a projected Hamiltonian $H_{\mathrm{circ,proj}}$ from the circuit reference model. We first identify the four eigenvectors of $H_{\mathrm{circ}}$ with the largest overlap with the bare states $\{\ket{q_1,0,q_0}\}_{q_1,q_0\in\{0,1\}}$. We then restrict these vectors to their computational components and apply Löwdin symmetric orthogonalization~\cite{lowdin_nonorthogonality_1970} to obtain an orthonormal basis that remains as close as possible to the original vectors. The projected Hamiltonian $H_{\mathrm{circ,proj}}$ is the diagonal matrix of the corresponding eigenvalues expressed in this basis.

The reduced models are calibrated against static quantities extracted from $H_{\mathrm{circ,proj}}$. For the effective model, the flux-dependent parameters $\tilde{\omega}_j(\phi)$, $J(\phi)$, and $\zeta(\phi)$ are fitted directly from the projected spectrum. For the Duffing model, we first obtain pointwise estimates of $\omega_j(\phi)$ and $\alpha_j(\phi)$ from \texttt{scqubits}, then refine them so that the resulting Duffing Hamiltonian matches the static quantities $J$ and $\zeta$ as closely as possible, and finally compress the flux dependence of each parameter with a low-order harmonic fit.\footnote{\url{https://github.com/SPS-Lab/quantum-circuit-modelling}}

For time-domain benchmarks, we solve the time-dependent Schr\"odinger equation under a prescribed control pulse: a drive pulse for the $R_X$ benchmark and a flux pulse for the CZ benchmark. The CZ target flux is chosen from the static analysis of the reference circuit model, in a region where the residual interaction $\zeta$ is large in magnitude. To isolate model differences under the same physical control, we apply the same pulse schedule to all three models. Each pulse is discretized on a time grid and approximated as piecewise constant, yielding a sequence of Hamiltonians $\{H[k]\}_{k=0}^{K-1}$. Starting from an initial state $\psi_k$, we propagate one step by
\begin{equation}
U_k=\mathrm{expm}\!\bigl(-i\,2\pi\,dt\,H[k]\bigr), \qquad
\psi_{k+1}=U_k\psi_k ,
\end{equation}
where \texttt{expm} is the matrix-exponential routine from \texttt{SciPy}. At each step, we record the amplitudes on the computational subspace and derive the reported observables from these stored trajectories. For a discretization with $K$ time steps, this procedure requires $K$ matrix exponentials and $K$ matrix-vector products. The runtime therefore scales primarily with the Hamiltonian dimension: $N_{E,q}^2N_{E,c}$ for the circuit and Duffing models, and $4$ for the effective model, since it acts only on the four-dimensional computational subspace. The dimension used in the benchmarks is selected from the plateau of a truncation convergence test. Specifically, we set the circuit dimensions to $N_Q=23, N_{E,q}=9, N_{E,c}=6$, and the Duffing dimensions to $N_{E,q}=N_{E,c}=3$.

\section{Results}
\begin{figure}[t]
  \centering
  \includegraphics{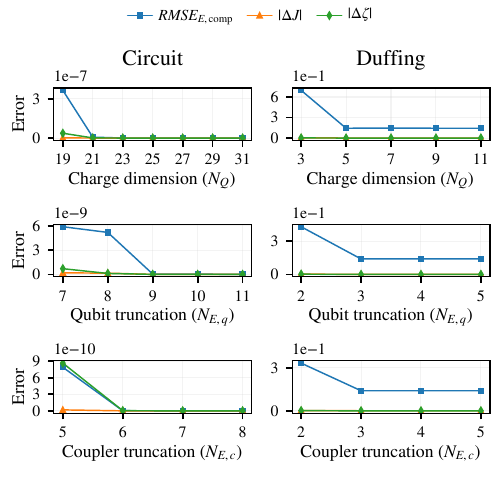}
  \caption{Truncation convergence of the circuit-based model (left) and Duffing model (right). The panels show the extracted quantities $J$ and $\zeta$, and the RMSE of the computational energy spectrum. Top: dependence on the charge-basis dimension $N_Q$ used for each transmon (extraction of initial guess of $\omega_{j}, \alpha_j$ in the Duffing). Middle: dependence on the retained qubit energy-basis dimension $N_{E,q}$. Bottom: dependence on the retained coupler dimension $N_{E,c}$. Errors are reported for the spectral RMSE and the extracted quantities $J$ and $\zeta$ relative to a highly resolved reference calculation.}
  \label{fig:truncation_benchmark}
\end{figure}

\begin{figure}
    \centering
    \includegraphics{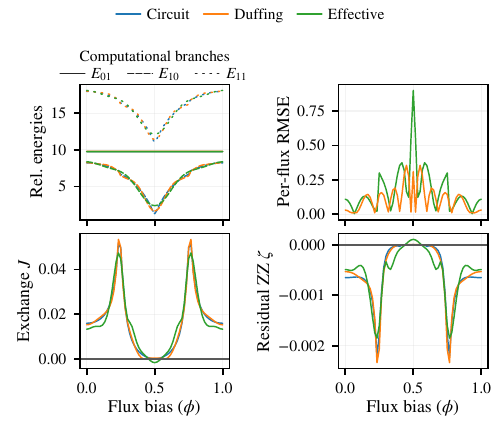}
    \caption{Static comparison across flux on qubit $q_1$. The panels show dressed computational energies relative to the ground state, per-flux spectral RMSE with respect to the circuit-based model, the extracted exchange-like coupling $J(\phi)$, and the residual interaction $\zeta(\phi)$.}
    \label{fig:static}
\end{figure}

\noindent\textbf{Reference convergence and static comparison across flux.}
Before comparing the reduced models, we first verify that the circuit-based model is sufficiently converged with respect to basis truncation. Figure~\ref{fig:truncation_benchmark} reports the dependence of the static error metrics on the charge-basis dimension $N_Q$, the retained qubit energy-basis dimension $N_{E,q}$, and the retained coupler dimension $N_{E,c}$. The errors are measured relative to a highly resolved reference calculation and as an average over a representative set of flux points. Across all three truncation axes, the spectral RMSE, $|dJ|$, and $|d\zeta|$ decrease rapidly and then level off, indicating that the production truncation lies in a numerically stable regime.
With the circuit-based reference fixed, we compare the three Hamiltonian descriptions across flux. Figure~\ref{fig:static} shows the dressed computational energies, the per-flux spectral RMSE relative to the circuit-based model, and the extracted quantities $J(\phi)$ and $\zeta(\phi)$. The state assignment in this figure is based on bare-state overlap at each flux point.
Three conclusions are immediate. First, the effective two-level model captures the broad low-energy trends, but its error grows in regions where the flux dependence becomes stronger, especially for the extracted quantities $J(\phi)$ and $\zeta(\phi)$. Second, the Duffing model follows the circuit-based model more closely over a larger portion of the flux range, both in the dressed energies and in the reduced two-qubit quantities. Third, the disagreement is strongly operating-point dependent: agreement near one flux value does not imply agreement across the full range.

To separate model-form error from truncation error in the Duffing description, we perform a second truncation study in which the errors are averaged over the full flux range. Figure~\ref{fig:truncation_benchmark} shows how the Duffing energies and extracted quantities approach the circuit-based reference as the Duffing qubit truncation is increased. The low-energy spectrum and the derived static metrics converge systematically, showing that the remaining discrepancy in Figure~\ref{fig:static} is not explained solely by insufficient Duffing truncation. The production truncations for the other benchmarks are chosen in the plateau region of all three metrics.\\

\begin{figure}
    \centering
    \includegraphics{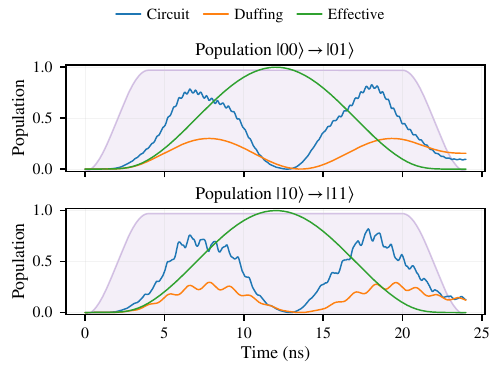}
    \caption{$R_X$ benchmark on $q_0$ with spectator qubit $q_1$ initialized in $\ket{0}$ (top) and $\ket{1}$ (bottom). The curves show the population transfer within the computational subspace predicted by the three models under the same drive pulse.}
    \label{fig:rx_populations}
\end{figure}

\noindent \textbf{Driven single-qubit dynamics.} We next evaluate the models on a time-domain single-qubit control task. An $R_X$-type pulse is applied to $q_0$, so $q_0$ is the \emph{driven qubit}. The second qubit, $q_1$, is not driven directly, but it remains coupled to the system through the shared bus. We therefore refer to $q_1$ as the \emph{spectator qubit}: it is present during the operation and can influence the pulse response even though it is not the target of the control. To test whether the models treat the driven qubit independently of the spectator state, we apply the same pulse twice, once with $q_1$ initialized in $\ket{0}$ and once with $q_1$ initialized in $\ket{1}$. Figure~\ref{fig:rx_populations} shows the resulting population transfer within the computational subspace. The two panels correspond to the transitions $\ket{00}\rightarrow\ket{01}$ and $\ket{10}\rightarrow\ket{11}$, respectively.

This benchmark already separates the models clearly. The effective model produces smooth, nearly ideal transfer because its state space is restricted to the computational subspace and therefore excludes multilevel distortions by construction. The Duffing and circuit models instead show oscillatory distortions and reduced transfer fidelity, indicating that the same pulse is coupling not only the intended computational transition but also additional degrees of freedom associated with higher qubit levels and the shared bus.

\begin{figure}
  \centering
  \includegraphics{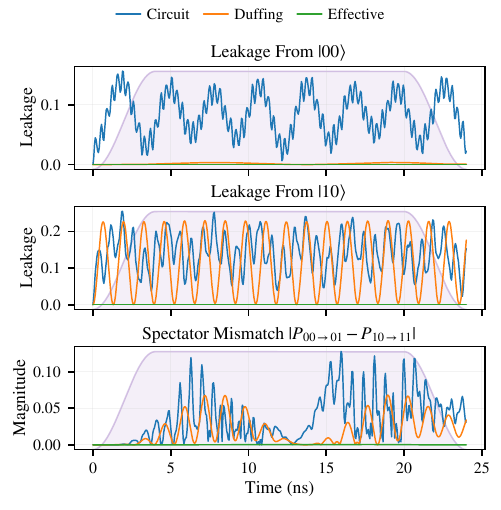}
  \caption{Diagnostics for the driven $R_X$ benchmark. Top and middle: leakage for the two spectator initializations. Bottom: spectator-state mismatch, defined as the absolute difference between the population transfer curves for $\ket{00}\rightarrow\ket{01}$ and $\ket{10}\rightarrow\ket{11}$.}
  \label{fig:rx_diagnostic}
\end{figure}

\begin{figure}
    \centering
    \includegraphics{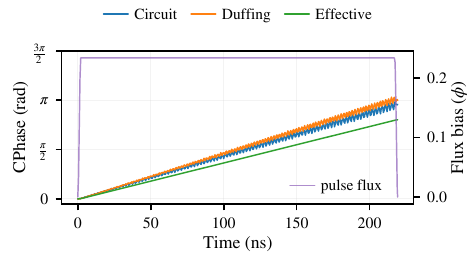}
    \caption{CZ gate dynamics under a flux pulse applied to $q_1$. The purple curve shows the reduced flux bias, and the colored curves show the accumulated conditional phase $\phi_{11}+\phi_{00}-\phi_{01}-\phi_{10}$ for an initial state $\ket{++}$ in the effective computational basis.}
    \label{fig:cz_dynamics}
\end{figure}

These differences become clearer in Figure~\ref{fig:rx_diagnostic}, which reports leakage and a spectator-state mismatch diagnostic. The mismatch is defined as
$|P_{\ket{00}\rightarrow\ket{01}}-P_{\ket{10}\rightarrow\ket{11}}|$,
that is, the absolute difference between the two transfer curves. It measures how much the pulse response of the driven qubit depends on the initial state of the spectator qubit, and would vanish for an ideal spectator-independent single-qubit rotation. The effective model predicts negligible mismatch and essentially no leakage. By contrast, the circuit model shows clear spectator dependence and substantial leakage, especially when the spectator is initialized in $\ket{1}$. The Duffing model reproduces the same qualitative pattern, although it does not match the full amplitude or fine temporal structure of the circuit model. \\
\noindent \textbf{CZ gate dynamics.} We now turn to the two-qubit time-domain benchmark. In this experiment, a flux pulse is applied to qubit $q_1$. The target flux is at $\phi=0.233$, in a region where the static analysis predicts a negative peak in the residual interaction $\zeta(\phi)$, so that the pulse induces conditional phase accumulation between the two qubits. The pulse duration is chosen as $t_{\mathrm{CZ}}=1/(2|\zeta|)$, with a $2\,\mathrm{ns}$ ramp on and off, so that the resulting evolution implements a CZ gate up to global and single-qubit phases. Figure~\ref{fig:cz_dynamics} shows both the applied flux pulse and the resulting conditional phase $\phi_{11}+\phi_{00}-\phi_{01}-\phi_{10}$. All three models predict approximately linear phase accumulation over the flat part of the pulse, but the detailed rates differ slightly. For this observable, the effective model stays close to the curves of the more detailed models. This is expected because the effective model is parameterized directly in terms of the extracted $ZZ$ interaction, which is the reduced quantity that controls conditional phase accumulation in the two-level description.

\begin{figure}
    \centering
    \includegraphics{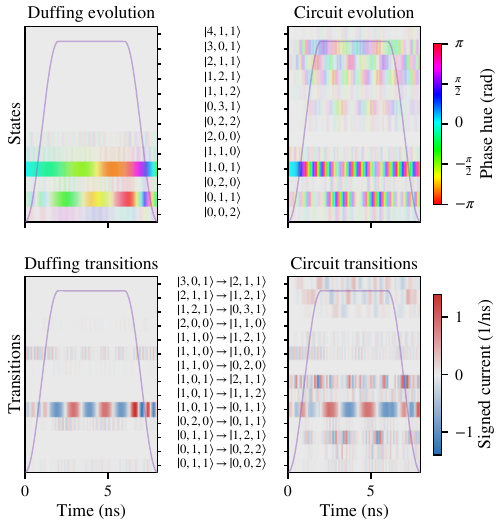}
    \caption{Leakage pathways during the CZ pulse for an initial state $\ket{1,0,1}$, corresponding to the computational state $\ket{11}$. Top row: state populations with phase encoded by color hue. Bottom row: signed population currents between the dominant states. States are ordered first by total excitation number and then lexicographically. The effective model is omitted because it does not contain explicit non-computational states.}
    \label{fig:leakage}
\end{figure}

\noindent \textbf{Leakage pathways during the CZ pulse.} The conditional phase alone does not reveal the full structure of the driven two-qubit evolution. Therefore, we examine leakage outside the computational subspace during the CZ pulse. To resolve the dominant leakage channels more clearly, we consider a shorter time window and initialize the system in $\ket{1,0,1}$, corresponding to the computational state $\ket{11}$. Figure~\ref{fig:leakage} shows the state populations and population currents for the Duffing and circuit-based models. The effective model is omitted because it has no explicit non-computational levels and therefore cannot represent leakage channels directly. In both the Duffing and circuit models, the dominant leakage path out of $\ket{1,0,1}$ leads first toward $\ket{0,1,1}$, indicating temporary transfer of excitation from qubit $q_1$ into the bus mode as resonance is approached. The Duffing model also captures leakage into states such as $\ket{2,0,0}$, showing that it already resolves a nontrivial leakage network. The circuit model, however, populates a broader set of states and reveals additional channels that are absent or strongly suppressed in the Duffing description.
The difference between the Duffing and circuit models in Figure~\ref{fig:leakage} is important for interpretation. The Duffing model is sufficient to detect that the CZ pulse is leakage-prone and to identify the main low-lying leakage paths. The circuit model is needed, however, to resolve the fuller structure of the leakage flow and the participation of higher excited states. This is precisely the type of distinction that is invisible in the effective two-level description. \\

\begin{figure}
    \centering
    \includegraphics{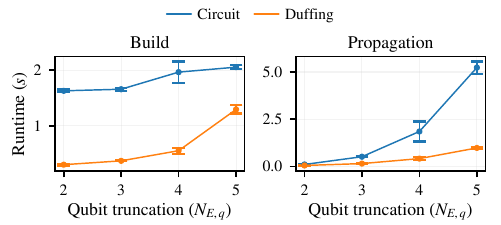}
    \caption{Runtime of the CZ simulation as a function of qubit truncation for the Duffing and circuit-based models. Left: Hamiltonian build time for the flux-point stack. Right: state-propagation time using the resulting Hamiltonians.}
    \label{fig:runtime}
\end{figure}

\noindent \textbf{Computational cost.} Finally, we compare the numerical cost of the Duffing and circuit-based models in the CZ simulation. We run the benchmark on an Acer Swift SF313-52G with i7-1065G7 CPU (4 cores) and 16GB RAM. Figure~\ref{fig:runtime} shows both the time required to build the stack of Hamiltonians over the sampled flux points and the time required to propagate the state, as functions of the qubit truncation.
The two models show the same qualitative trend: increasing truncation increases both build time and propagation time. However, the circuit-based model is consistently more expensive, and the propagation cost grows more steeply with truncation than in the Duffing model. This is consistent with the larger basis construction and denser multilevel structure of the circuit model.

\section{Discussion and Conclusion}
We compared three Hamiltonian descriptions for pulse-level simulation of a flux-tunable two-qubit system with a coupler. Our results show that the effective model is useful as a compact reduced description for static observables, while the Duffing model offers a practical middle ground by reproducing flux-dependent behavior and low-lying multilevel effects more faithfully at lower cost than the circuit-based description. However, for $R_X$ gate dynamics and complex leakage flows during CZ dynamics, richer multilevel models are necessary, since static agreement alone does not ensure accurate time-domain behavior.

From a practical standpoint, this suggests that circuit-based models should primarily be used as high-fidelity references for calibrating and validating small subsystems, while Duffing-type models are the more realistic default for pulse-level simulation of larger circuits. Effective two-level models remain useful for tasks with reduced complexity confined to the computational subspace, but are too coarse when multilevel effects must be resolved, such as observing leakage in microwave-driven pulses.

As future work, it would be natural to extend the comparison to additional driven-control settings, including optimized two-qubit pulses, to test whether the conclusions found here persist beyond the flux-pulsed CZ case and the single-qubit $R_X$ case. It would also be valuable to study larger multi-qubit devices, where the balance between physical fidelity and model reduction becomes even more important for practical pulse-level simulation.

\begin{acks}
This work was supported by the Swedish Wallenberg Centre for Quantum Technology (WACQT).
\end{acks}

\bibliographystyle{ACM-Reference-Format}
\bibliography{main}

\clearpage
\end{document}